\documentclass[11pt]{article}
\usepackage{amssymb,cite}
\makeatletter
\@addtoreset{equation}{section}
\makeatother

\def\ben{\begin{equation}}
\def\een{\end{equation}}

 \def\bd{\begin{document}} \def\ed{\end{document}}
\def\ds{\documentstyle} \let\fr=\frac \let\bl=\bigl \let\br=\bigr
\let\Br=\Bigr \let\Bl=\Bigl
\let\bm=\bibitem
\let\na=\nabla
\let\pa=\partial \let\ov=\overline
\newcommand{\be}{\begin{equation}}
\newcommand{\ee}{\end{equation}}
\def\ba{\begin{array}}
\def\ea{\end{array}}
\def\ft#1#2{{\textstyle{{\scriptstyle #1}\over {\scriptstyle #2}}}}
\def\fft#1#2{{#1 \over #2}}
\def\del{\partial}
\def\vp{\varphi}
\def\sst#1{{\scriptscriptstyle #1}}
\def\oneone{\rlap 1\mkern4mu{\rm l}}
\def\td{\tilde}
\def\wtd{\widetilde}
\def\ie{\rm i.e.\ }
\def\dalemb#1#2{{\vbox{\hrule height .#2pt
        \hbox{\vrule width.#2pt height#1pt \kern#1pt
                \vrule width.#2pt}
        \hrule height.#2pt}}}
\def\square{\mathord{\dalemb{6.8}{7}\hbox{\hskip1pt}}}
\newcommand{\ho}[1]{$\, ^{#1}$}
\newcommand{\hoch}[1]{$\, ^{#1}$}
\newcommand{\bea}{\begin{eqnarray}}
\newcommand{\eea}{\end{eqnarray}}

\title{Integrability of Some Charged Rotating Black Hole Solutions of Supergravity}

\newcommand{\auth}{Muraari Vasudevan}

\begin{document}

\begin{flushright}

Alberta Thy 09-05\\
PACS numbers: 04.50.+h, 98.80.Cq\hfill\\
August\  2005
\end{flushright}

\vspace{10pt}

\begin{center}
{\large {\bf Integrability of Some Charged Rotating Supergravity
Black Hole Solutions in Four and Five Dimensions
            }}

\vspace{20pt}
\auth

\vspace{10pt}

{\it  Theoretical Physics Institute, University of Alberta,\\
Edmonton, Alberta  T6G 2J1, Canada}

{\it
    {\rm E-mail: \texttt{mvasudev@phys.ualberta.ca}}
  }


\vspace{40pt}

\underline{ABSTRACT}
\end{center}

We study the integrability of geodesic flow in the background of
some recently discovered charged rotating solutions of
supergravity in four and five dimensions. Specifically, we work
with the gauged multicharge Taub-NUT-Kerr-(Anti) de Sitter metric
in four dimensions, and the $U(1)^3$ gauged charged-Kerr-(Anti) de
Sitter black hole solution of N = 2 supergravity in five
dimensions. We explicitly construct the Killing tensors that
permit separation of the Hamilton-Jacobi equation in these
spacetimes. These results prove integrability for a large class of
previously known supergravity solutions, including several BPS
solitonic states. We also derive first-order equations of motion
for particles in these backgrounds and examine some of their
properties. Finally, we also examine the Klein-Gordon equation for
a scalar field in these spacetimes and demonstrate separability.

\pagebreak

\section{Introduction}

Solutions of the vacuum Einstein equations describing black hole
solutions in both four and higher dimensions are currently of
great interest. This is mainly due to a number of recent
developments in high energy physics. Models of spacetimes with
large extra dimensions have been proposed to deal with several
questions arising in modern particle phenomenology (eat about.g.
the hierarchy problem) \cite{NDD, ANDD, RS}. These models allow
for the existence of higher dimensional black holes which can be
described classically. Also of interest in these models is the
possibility of mini black hole production in high energy particle
colliders which, if they occur, provide a window into
non-perturbative gravitational physics \cite{Cav, Kan}.

Superstring and M-Theory, which call for additional spacetime
dimensions, naturally incorporate black hole solutions in higher
dimensions (10 or 11). P-branes present in these theories can also
support black holes, thereby making black hole solutions in an
intermediate number of dimensions physically interesting as well.
Black hole solutions in superstring theory are particularly
relevant since they can be described as solitonic objects. They
provide important keys to understanding strongly coupled
non-perturbative phenomena which cannot be ignored at the
Planck/string scale \cite{DV, CT}.

Astrophysically relevant black hole spacetimes are, to a very good
approximation, described by the Kerr metric \cite{Kerr}. One
generalization of the Kerr metric to higher dimensions is given by
the Myers-Perry construction \cite{MyersPerry}. With interest now
in a nonzero cosmological constant, it is worth studying
spacetimes describing rotating black holes with a cosmological
constant. Another motivation for including a cosmological constant
is driven by the AdS/CFT correspondence. The study of black holes
in an Anti-de Sitter background could give rise to interesting
descriptions in terms of the conformal field theory on the
boundary leading to better understanding of the correspondence
\cite{Mal, Wit}. The general Kerr-de Sitter metrics describing
rotating black holes in the presence of a cosmological constant
have been constructed explicitly in \cite{GPLP, GPLP2}.

There is a strong need to understand explicitly the structure of
geodesics in the background of black holes in Anti-de Sitter space
in the context of string theory and the AdS/CFT correspondence.
This is due to the recent work in exploring black hole singularity
structure using geodesics and correlators in the dual CFT on the
boundary \cite{Shen1, Shen2, Shen3, Shen4, Shen5, Shen6}. The
metrics mentioned above, have so far proven to yield little or no
information through an analysis of this sort. Black holes with
charge are particularly interesting for this type of analysis
since the charges are reinterpreted as the R-charges of the dual
theory. The spacetimes explored in this paper are exact solutions
of supergravity in backgrounds with a cosmological constant and
charges, and thus could be more suitable for this sort of geodesic
analysis.

In this paper we work with the four-dimensional multicharge
Kerr-Taub-NUT-(Anti) de Sitter solution of supergravity recently
discovered by Chong, Cvetic, Lu, and Pope \cite{CCLP}, as well as
the $U(1)^3$ gauged Kerr-(Anti) de Sitter black hole solution of N
= 2 supergravity in five dimensions discovered by Cvetic, Lu, and
Pope\cite{CLP}.

We study the separability of the Hamilton-Jacobi equation in these
spacetimes, which can be used to describe the motion of classical
massive and massless particles (including photons). We use this
explicit separation to obtain first-order equations of motion for
both massive and massless particles in these backgrounds. The
equations are obtained in a form  that could be used for numerical
study, and also in the study of black hole singularity structure
using geodesic probes and the AdS/CFT correspondence.

We also study the Klein-Gordon equation describing the propagation
of a massive scalar field in these spacetimes. Separation again
turns out to be possible with the usual multiplicative ansatz.

This paper greatly generalizes the results of \cite{frolov1 ,
frolov2} for the Myers-Perry metric in five dimensions,
\cite{VSP1} which separates the equations in the case of equal
rotation parameters in the odd dimensional Kerr-(A) dS spacetimes,
\cite{KL} which separates the equations for the general five
dimensional Kerr-(A) dS spacetime with unequal rotation
parameters, \cite{VSP2} which separates the equations in the case
of two independent sets of rotation parameters in the Myers-Perry
metrics in all dimensions, \cite{VS1} which separates the
equations in the case of two independent sets of rotation
parameters in the Kerr-(A) dS metrics in all dimensions, and
\cite{ZGLP} which separates the equations in the case of a single
non-zero rotation parameter for uncharged Kerr-Taub-NUT metrics in
arbitrary dimensions. Some further work for other special cases
were also done in \cite{MDS1} and \cite{KL2}.

Separation turns out to be possible for both equations in these
metrics due to the existence of second-order Killing tensors, one
of them non-trivial and irreducible. This is a generalization of
the Killing tensor in the Kerr black hole spacetime in four
dimensions constructed in \cite{Carter} which was subsequently
described by Chandrasekhar as the ``miraculous property of the
Kerr metric".  A similar construction for the Myers-Perry metrics
in higher dimensions has also been done\cite{frolov1, VSP2}, and
for the Kerr-Taub-NUT metrics in arbitrary dimensions without
charge and only one nonzero rotation parameter in \cite{ZGLP}. The
Killing tensors, in each case, provides an additional integral of
motion necessary for complete integrability.

\section{Overview of the Metrics}
\subsection{Four Dimensional Kerr-Taub-NUT Multicharge Gauged Solution of
Supergravity}

This metric was recently obtained by Chong, Cvetic, Lu, and Pope
in \cite{CCLP}. The solution was obtained by starting out with the
four dimensional Kerr-Taub-NUT metric, dimensionally reducing to
three dimensions along the time direction, and then lifting back
up after ``dualizing". The metric is given by
\begin{eqnarray}
ds^2=-\frac{\Delta_r}{a^2W}[adt + u_1 u_2 d\phi]^2 +
\frac{\Delta_u}{a^2W}[adt-r_1r_2d\phi]^2+W\left[\frac{dr^2}{\Delta_r}+
\frac{du^2}{\Delta_u}\right] \, , \label{nutmet}
\end{eqnarray}
where
\begin{eqnarray}
W&=&r_1r_2+u_1u_2 \, , \quad r_i=r+2ms_i^2 \, , \quad u_i = u +2ls_i ^2 \, ,\quad i=1,2\, , \nonumber \\
\Delta_r &=& r^2 + a^2 -2mr + g^2 r_1r_2(r_1r_2+a^2) \, ,\nonumber\\
\Delta_u &=& -u^2+a^2 +2lu+g^2u_1u_2(u_1u_2-a^2) \, ,
\end{eqnarray}
and we use the notation
\begin{eqnarray}
s_i=\sinh \delta _i \, , \quad c_i=\cosh \delta _i \, , \quad
i=1,2\, .
\end{eqnarray}
Here $\delta_1$ is the magnetic charge, $\delta_2$ is the electric
charge, $l$ is the NUT parameter, $a$ is the rotation parameter,
and $g$ is the gauge parameter. The cosmological constant
$\Lambda$ is given by $\Lambda=-g^2$. The ungauged solution is
obtained by setting $g$ to zero.

If the two charge parameters are set equal, $\delta_1=\delta_2$,
then the solution reduces to the charged AdS-Kerr-Taub-NUT
solution of Einstein-Maxwell theory with a cosmological constant.
To reduce to the usual coordinate system, we use the change of
coordinate $u=a\cos\theta$. With $l$ set to zero, we recover the
metric found in \cite{CCLP} for a multicharge Kerr-(Anti) de
Sitter black hole in gauged supergravity in four dimensions.

For future reference, we note the following expressions. The
determinant of the metric is given by
\begin{eqnarray}
g=-\frac{W^2}{a} \, . \label{detnut}
\end{eqnarray}

The components of the inverse metric are
\begin{eqnarray}
g^{tt}&=&\frac{1}{\Delta_r\Delta_u W}[\Delta_r u_1^2
u_2^2-\Delta_u r_1^2 r_2 ^2] \, , \quad g^{\phi
\phi}=\frac{a^2}{\Delta_r\Delta_u
W}[\Delta_r-\Delta_u]\, , \nonumber \\
g^{t\phi}&=&\frac{a}{\Delta_r\Delta_u W} [\Delta_r u_1u_2+\Delta_u
r_1r_2] \, , \quad g^{rr}=\frac{\Delta_r}{W} \, , \quad
g^{uu}=\frac{\Delta_u}{W} \label{invnut} \, .
\end{eqnarray}
We also note that the functions $\Delta_r$ and $\Delta_u$ are
functions of $r$ and $u$ only, respectively.

\subsection{$U(1)^3$ Gauged Kerr-(Anti) de Sitter
Black Hole Solution of $\mathcal{N} = 2$ Supergravity in Five
Dimensions}

This metric was recently obtained by Chong, Lu, and Pope in
\cite{CLP}. The metric is given by
\begin{eqnarray}
ds^2=&-&\frac{Y-f_3}{R^2}dt^2+\frac{r^2R}{Y}dr^2+ Rd\Omega_3^2 +
\frac{f_1-R^3}{R^2}(\sin^2 \theta d\phi+ \cos ^2 \theta d\psi)^2
\nonumber\\
&-& \frac{2f_2}{R^2}dt(\sin^2\theta d\phi + \cos ^2 \theta d\psi)
\, ,
\end{eqnarray}
where
\begin{eqnarray}
R&=&r^2\left(\prod_{i=1}^3 H_i\right)^{1/3} \, , \quad
H_i=1+\frac{Ms_i^2}{r^2} \, , \nonumber \\
d\Omega_3^2 &=& d\theta ^2 + sin\theta ^2 d\phi ^2 + \cos ^2
\theta d\psi ^2 \, ,
\end{eqnarray}
and as before
\begin{eqnarray}
s_i=\sinh \delta _i \, , \quad c_i=\cosh \delta _i \, , \quad
i=1,2,3\, .
\end{eqnarray}
The functions $f_i$, and $Y$ are defined by
\begin{eqnarray}
f_1&=&R^3 + Ma^2 r^2 + M^2a^2\left[2\left(\prod_i c_i - \prod _i
s_i\right) \prod_j s_j-\sum_{i<j}s_i^2 s_j^2\right] \, ,
\nonumber \\
f_2&=& \gamma a\Lambda R^3+ Ma\left(\prod_i c_i -
\prod_i s_i\right)r^2 + M^2 a \prod_i s_i \, , \nonumber \\
f_3&=& \gamma ^2 a^2 \Lambda ^2 R^3 + M a^2 \Lambda
\left[2\gamma\left(\prod_i c_i - \prod_i s_i\right)-\Sigma\right]
r^2 \nonumber \\
&&\, + Ma^2 -\Lambda \Sigma M^2 a^2 \left[2(\left(\prod_i c_i -
\prod_i s_i\right)\prod_j s_j - \sum_{i<j} s_i^2 s_j^2\right] +
2\lambda \gamma M^2 a^2 \prod_i s_i \, , \nonumber \\
Y&=& f_3-\Lambda \Sigma R^3 + r^4 - Mr^2 \, ,
\end{eqnarray}
and
\begin{eqnarray}
\Sigma=1+\gamma ^2 a^2 \Lambda.
\end{eqnarray}
It is important to note that these are functions of the coordinate
$r$ only.

The parameter $M$ is related to the mass of the black hole, the
$\delta _i$ are the charges associated with each of the three
$U(1)$ gauge groups, the gauge parameter $g$ is related to the
cosmological constant $\Lambda$ via $\Lambda=-g^2$, $a$ is the
rotation parameter of the black hole (equal rotation parameters in
the two independent planes was assumed in the derivation of the
metric), and the constant $\gamma$ is simply a redundant parameter
which is useful to test several limits, but could be eliminated if
necessary. This metric encompasses, as special limits, several
previously known solutions such as the Klemm-Sabra BPS solution
etc. More details about these limits can be found in \cite{CCLP}.

In order to avoid long complicated expressions, we introduce the
following functions to write the metric more compactly
\begin{eqnarray}
A(r)=\frac{Y-f_3}{R^2}\, , \quad W(r)=\frac{Y}{r^2 R}\, , \quad
B(r)=\frac{f_1-R^3}{R^2}\, , \quad C(r)=-\frac{f_2}{R^2} \, .
\end{eqnarray}
Note that all of these are functions of the coordinate $r$ only.
The metric is then written compactly in the form
\begin{eqnarray}
ds^2=&-&A(r)dt^2 + \frac{dr^2}{W(r)}+R d\Omega_3^2 +
B(r)(\sin^2\theta d\phi+\cos^2\theta d\psi)^2 \nonumber \\
&+& 2C(r)dt(\sin ^2 \theta d\phi + \cos ^2 \theta d\psi) \, .
\end{eqnarray}

The components of the inverse metric are
\begin{eqnarray}
g^{rr}&=&W(r)\, , \nonumber \\
g^{\theta \theta}&=& \frac{1}{R} \, , \nonumber \\
g^{tt}&=&-\frac{B(r)+R}{r^2W(r)} \, , \nonumber \\
g^{t\phi}&=&g^{t\psi}=\frac{C(r)}{r^2W(r)} \, ,\nonumber \\
g^{\phi\phi}&=&\frac{A(r)B(r)\cos^2\theta +
A(r)R+C^2(r)\cos^2\theta}{Rr^2W(r)\sin^2\theta} \,, \nonumber \\
g^{\psi\psi}&=&\frac{A(r)B(r)\sin^2\theta +
A(r)R+C^2(r)\sin^2\theta}{Rr^2W(r)\cos^2\theta} \,, \nonumber \\
g^{\phi\psi}&=&-\frac{A(r)B(r)+C^2(r)}{Rr^2W(r)} \, .
\label{invu1}
\end{eqnarray}

We note for future reference the following identity which can
easily be verified using Maple \cite{Map}
\begin{eqnarray}
A(r)B(r)+A(r)R+C^2(r)=r^2W(r) \, .
\end{eqnarray}
Finally, the determinant of the metric can be calculated to be
\begin{eqnarray}
g=-r^2R^2\sin^2\theta \cos^2\theta \, , \label{detu1}
\end{eqnarray}
where we need to make use of the identity given above repeatedly.

\section{Integrals of Motion and the Hamilton-Jacobi Equation}
The equations of motion of a test particle of mass $m$ in a
gravitational background described by a metric $g_{\mu \nu}$ are
\begin{eqnarray}
\frac{D^2x^{\mu}}{D\tau^2}=0 \, ,
\end{eqnarray}
where $\frac{D}{D\tau}$ is the covariant derivative with respect
to proper time $\tau$. These equations can be derived from a
Lagrangian
\begin{eqnarray}
L=\frac{1}{2}g_{\mu \nu} \dot{x^\mu}\dot{x^\nu} \, ,
\end{eqnarray}
where a dot denotes a partial derivative with respect to an affine
parameter $\lambda$. This can be chosen such that $\tau=m\lambda$.

The symmetries of the metric, if any, can provide us with some
integrals of motion. For instance, if the metric is stationary,
i.e. does not depend on the time $t$, then the energy is
conserved. However, in most situations, sufficient number of
integrals of motion do not exist. Also, using the Lagrangian
formulation, sometimes certain integrals of motion are impossible
to obtain even if they exist. Usually these are ``second order" in
the momenta such as the case of the Carter constant for the Kerr
metric. Such additional integral of motion, which permit us, in
these cases to completely integrate the equations of motion, can
be provided by the Hamilton-Jacobi equation (though a proper
choice of coordinate system is necessary).

The Hamilton-Jacobi equation in a curved background is given by
\be -\frac{\partial S}{\partial \lambda} = H = \frac{1}{2} g^{\mu
\nu } \frac{\partial S}{\partial x^{\mu}} \frac{\partial
S}{\partial x^{\nu}} \,, \label{HJ} \ee where $S$ is the action
associated with the particle and $\lambda$ is some affine
parameter along the worldline of the particle. Note that this
treatment also accommodates the case of massless particles, where
the trajectory cannot be parametrized by proper time.

\section{Particle Motion in the Four Dimensional Kerr-Taub-NUT Multicharge Gauged
Solution of Supergravity}

\subsection{Separation of Variables}
We can attempt a separation of coordinates as follows. Let
\begin{equation}
S=\frac{1}{2}m^2 \lambda -Et + L_\phi \phi + S_\theta (\theta) +
S_r (r)\,. \label{ansatznut}
\end{equation}
$t$ and $\phi$ are cyclic coordinates, so their conjugate momenta
are conserved. The conserved quantity associated with time
translation is the energy $E$, and that with rotation in $\phi$ is
the corresponding angular momentum $L_\phi$. Then using the
components of the inverse metric (\ref{invnut}), the
Hamilton-Jacobi equation (\ref{HJ}) is written to be
\begin{eqnarray}
-m^2&=&\frac{1}{\Delta_r \Delta_u W} [\Delta_r u_1^2 u_2^2 -
\Delta_u r_1^2 r_2^2](-E)^2 + \frac{a^2}{\Delta_r \Delta_u
W}[\Delta_r-\Delta_u]L_\phi ^2 + \frac{\Delta_r}{W}
\left[\frac{dS_r(r)}{dr}\right]^2 \nonumber \\
&&+\frac{\Delta_u}{W}\left[\frac{dS_u(u)}{du}\right]^2 +
\frac{2a}{\Delta_r \Delta_u W}[\Delta_r u_1 u_2 + \Delta _u r_1
r_2](-E)L_\phi \, .
\end{eqnarray}

Now multiplying both sides by $W$, we can separate out the
equation in the form
\begin{eqnarray}
K&=&\Delta_r \left[\frac{dS_r(r)}{dr}\right]^2 -
\frac{1}{\Delta_r}[r_1r_2E+aL_\phi]^2 + m^2r_1r_2 \, , \nonumber\\
K&=&-\Delta_u\left[\frac{dS_u(u)}{du}\right]^2-\frac{1}{\Delta_u}[u_1u_2E-aL_\phi]^2-m^2u_1u_2
\, , \label{sepnut}
\end{eqnarray}
where $K$ is a constant of separation.

\subsection{The Equations of Motion}

To derive the equations of motion, we will write the separated
action $S$ from the Hamilton-Jacobi equation in the following
form: \bea S&=&\frac{1}{2}m^2 \lambda -Et +L_\phi \phi + \int ^r
\sqrt{\mathcal{R}(r')} dr' + \int ^ {u} \sqrt{U(u')}du' \, , \eea
where
\begin{eqnarray}
\Delta_r \mathcal{R}(r)&=&K+\frac{1}{\Delta_r}[r_1r_2E+aL_\phi]^2 - m^2r_1r_2 \, , \nonumber\\
\Delta_uU(u)&=&-K-\frac{1}{\Delta_u}[u_1u_2E-aL_\phi]^2-m^2u_1u_2\,.
\end{eqnarray}

To obtain the equations of motion, we differentiate $S$ with
respect to the parameters $m^2,K,E,L_\phi$ and set these
derivatives to equal other constants of motion. However, we can
set all these new constants of motion to zero (following from
freedom in choice of origin for the corresponding coordinates, or
alternatively by changing the constants of integration). Following
this procedure, we get the following equations of motion:
\vspace{.5cm}
\begin{eqnarray}
\frac{\pa S}{\pa m^2}&=&0 \Rightarrow \lambda = \int r_1r_2 \frac{dr}{\Delta_r\sqrt{\mathcal R}} + \int u_1u_2\frac{du}{\Delta_u\sqrt{U}} \, , \nonumber \\
\frac{\pa S}{\pa K}&=&0 \Rightarrow \int \frac{du}{\Delta_u\sqrt{U}} = \int \frac{dr}{\Delta_r\sqrt{\mathcal R}} \, , \nonumber \\
\frac{\pa S}{\pa L_\phi}&=&0 \Rightarrow \phi=\int (r_1r_2E+aL_\phi)\frac{dr}{\Delta_r^2\sqrt{\mathcal R}}+ \int (u_1u_2E-aL_\phi) \frac{du}{\Delta_u^2\sqrt U}  \, , \nonumber \\
\frac{\pa S}{\pa E}&=&0 \Rightarrow t=\int
r_1r_2(r_1r_2E+aL_\phi)\frac{dr}{\Delta_r^2\sqrt{\mathcal R}}-
\int u_1u_2(u_1u_2E-aL_\phi) \frac{du}{\Delta_u^2\sqrt
U}\,.\label{inteqs}
\end{eqnarray}
It is often more convenient to rewrite these in the form of
first-order differential equations obtained from (\ref{inteqs}) by
direct differentiation with respect to the affine parameter:
\begin{eqnarray}
W \frac{dr}{dl} &=& \Delta_r \sqrt{\mathcal R} \, , \nonumber \\
W \frac{d u}{dl} &=& \Delta_u \sqrt{U} \, , \nonumber \\
W\frac{d\phi}{dl}&=&\frac{r_1r_2E+aL_\phi}{\Delta_r}+\frac{u_1u_2E-aL_\phi}{\Delta_\theta}
\, , \nonumber \\
W\frac{dt}{dl}&=&\frac{r_1r_2(r_1r_2E+aL_\phi)}{\Delta_r}-\frac{u_1u_2(u_1u_2E-aL_\phi)}{\Delta_\theta}
\, . \label{eqns}
\end{eqnarray}

\subsection{Analysis of the Radial Equation}
The worldline of particles in the background considered above are
completely specified by the values of the conserved quantities
$E,L_\phi, K$, and by the initial values of the coordinates. We
will consider particle motion in the black hole exterior. Allowed
regions of particle motion necessarily need to have positive value
for the quantity $R$, owing to equation (\ref{eqns}). We determine
some of the possibilities of the allowed motion.

We will consider the motion of a particle in the black hole
exterior. Thus we can assume that $\Delta_r>0$ for large $r$. At
large $r$, the dominant contribution to $\mathcal R$, in the case
of $\Lambda =0$, is $E^2 -m^2$. Since $\Lambda=-g^2$, zero
cosmological constant corresponds to the charged rotating black
hole in four dimensions in ungauged supergravity. Here, we can
thus say that for $E^2<m^2$, we cannot have unbounded orbits,
whereas for $E^2>m^2$, such orbits are possible. For the case of
nonzero $\Lambda$ (i.e. also nonzero $g$ which implies we are now
considering gauged supergravity), the dominant term at large $r$
in $R$ (or rather the slowest decaying term) is $-m^2r^2$. Thus in
the case of the Anti-de Sitter background (since $\Lambda=-g^2$ is
negative), only bound orbits are possible.

In order to study the radial motion of particles in these metrics,
it is useful to cast the radial equation of motion into a
different form. Decompose $\mathcal R$ as a quadratic in $E$ as
follows:
\begin{equation}
\mathcal R=\alpha E^2-2\beta E + \gamma \,,
\end{equation}
where
\begin{eqnarray}
\alpha &=& \frac{r_1^2r_2^2}{\Delta_r^2} \, , \nonumber \\
\beta &=& -\frac{r_1r_2aL_\phi}{\Delta_r^2} \, ,\nonumber \\
\gamma&=&\frac{K-m^2r_1r_2}{\Delta_r}+\frac{a^2L_\phi^2}{\Delta_r^2}
\, .
\end{eqnarray}

The turning points for trajectories in the radial motion (defined
by the condition $\mathcal R=0$) are given by $E=V_{\pm}$ where
\be V_{\pm} = \frac{\beta \pm \sqrt{\beta ^2 -\alpha \gamma
}}{\alpha} \,. \ee These functions, called the effective
potentials \cite{frolov1}, determine allowed regions of motion. In
this form, the radial equation is much more suitable for detailed
numerical analysis for specific values of parameters.

\section{Particle Motion in the $U(1)^3$ Gauged Kerr-(Anti) de Sitter Black Hole
Solution of $\mathcal{N} = 2$ Supergravity in Five Dimensions}

We will only sketch the analysis of the separation of variables
here, since the procedure for deriving equations of motion etc. is
virtually identical to those of the metric above.

We can attempt a separation of coordinates as follows. Let
\begin{equation}
S=\frac{1}{2}m^2 \lambda -Et + L_\phi \phi +L_\psi \psi+ S_\theta
(\theta) + S_r (r)\,. \label{ansatzu1}
\end{equation}
$t$, $\phi$, and $\psi$ are cyclic coordinates, so their conjugate
momenta are conserved. The conserved quantity associated with time
translation is the energy $E$, and those with rotation in $\phi$
and $\psi$ are the corresponding angular momenta $L_\phi$ and
$L_\psi$. Then using the components of the inverse metric
(\ref{invu1}), the Hamilton-Jacobi equation (\ref{HJ}) is written
to be
\begin{eqnarray}
-m^2&=&\frac{B(r)+R}{r^2W(r)}(-E)^2 +
2\frac{C(r)}{r^2W(r)}(-E)(L_\phi)+
2\frac{C(r)}{r^2W(r)}(-E)(L_\psi) \nonumber \\
&+&\frac{A(r)B(r)\cos^2\theta +
A(r)R+C^2(r)\cos^2\theta}{Rr^2W(r)\sin^2\theta}
L_\phi^2+\frac{1}{R}\left[\frac{dS_\theta(\theta)}{d\theta}\right]^2+
\nonumber \\
&+&\frac{A(r)B(r)\sin^2\theta
A(r)R+C^2(r)\sin^2\theta}{Rr^2W(r)\cos^2\theta} L_\psi^2 + W(r)
\left[\frac{dS_r(r)}{dr}\right]^2 \, .
\end{eqnarray}
After some algebraic manipulation and using some trigonometric
identites we can write this as
\begin{eqnarray}
-m^2&=&W(r)
\left[\frac{dS_r(r)}{dr}\right]^2-2\frac{C(r)E}{r^2W(r)}(L_\phi+L_\psi)-\frac{B(r)+R}{r^2W(r)}E^2
\nonumber \\
&-&\frac{A(r)B(r)+C^2(r)}{Rr^2W(r)}(L_\phi+L_\psi)^2 +
\frac{1}{R}(\csc^2\theta L_\phi ^2 +\sec^2\theta L_\psi
^2)+\frac{1}{R}\left[\frac{dS_\theta(\theta)}{d\theta}\right]^2\,
.
\end{eqnarray}
In this form, the Hamilton-Jacobi equation can now be easily
separated to give
\begin{eqnarray}
-K&=&\left[\frac{dS_\theta(\theta)}{d\theta}\right]^2+\csc^2\theta
L_\phi ^2 +\sec^2\theta L_\psi ^2 \, ,\nonumber \\
K&=&Rm^2+W(r)R
\left[\frac{dS_r(r)}{dr}\right]^2-2\frac{C(r)RE}{r^2W(r)}(L_\phi+L_\psi)-\frac{B(r)R+R^2}{r^2W(r)}E^2
\nonumber \\
&-&\frac{A(r)B(r)+C^2(r)}{r^2W(r)}(L_\phi+L_\psi)^2 \, ,
\label{sepu1}
\end{eqnarray}
where $K$ is a constant of separation.

To derive the equations of motion, the separated action $S$ from
the Hamilton-Jacobi equation is more conveniently written, as
before, in the following form: \bea S&=&\frac{1}{2}m^2 \lambda -Et
+L_\phi \phi + L_\psi \psi+ \int ^r \sqrt{\mathcal{R}(r')} dr' +
\int ^ {\theta} \sqrt{\Theta(\theta')}d\theta' \, , \eea where
\begin{eqnarray}
RW(r)
\mathcal{R}(r)&=&K-Rm^2+2\frac{C(r)RE}{r^2W(r)}(L_\phi+L_\psi)+\frac{B(r)R+R^2}{r^2W(r)}E^2\nonumber \\
&+&\frac{A(r)B(r)+C^2(r)}{r^2W(r)}(L_\phi+L_\psi)^2 \, , \nonumber\\
\Theta(\theta)&=&-K-\csc^2\theta L_\phi ^2 -\sec^2\theta L_\psi^2
\, .
\end{eqnarray}

By following the same procedure as earlier, we can easily
establish first order equations of motion, and a radial effective
potential etc. Since the derivation is remarkably similar, we will
not reproduce the results here in the interests of being concise.

\section{Dynamical Symmetry}
The general class of metrics discussed here are stationary and
``axisymmetric"; i.e., $\partial / \partial t$ and $\partial /
\partial \phi$ (as well as $\partial /
\partial \psi$ in the five dimensional $U(1)^3$ case) are Killing vectors and have associated
conserved quantities, $-E$ and $L_\phi$ (and $L_\psi$). In general
if $\xi$ is a Killing vector, then $\xi ^{\mu} p_{\mu}$ is a
conserved quantity, where $p$ is the momentum of the particle.
Note that this quantity is first order in the momenta.

As mentioned earlier, the additional constant of motion $K$ which
allowed for complete integrability of the equations of motion is
not related to a Killing vector from a cyclic coordinate. This
constant is, rather, derived from a non-trivial irreducible
second-order Killing tensor in both spacetimes, which permits the
separation of the $r-\theta$ (or $r-u$) equations in both cases.
These Killing tensors are generalizations of the Killing tensor
obtained in four dimensions by Carter \cite{Carter} and in five
dimensions for the Myers-Perry metric in \cite{frolov1}. Killing
tensors are not symmetries on configuration space, and cannot be
derived from a Noether procedure, and are rather, symmetries on
phase space. They obey a generalization of the Killing equation
for Killing vectors (which do generate symmetries in configuration
space by the Noether procedure) given by
\begin{eqnarray}
K_{(\mu\nu;\rho)}=0 \, ,
\end{eqnarray}
where $K$ is any second order Killing tensor, and the parentheses
indicate complete symmetrization of all indices.

The Killing tensors can be obtained from the expressions for the
separation constant $K$ in each case. If the particle has momentum
$p$, then the Killing tensor $\mathcal K_{\mu \nu}$ is related to
the constant $K$ via
\begin{eqnarray}
K=\mathcal K^{\mu \nu}p_{\mu}p_{\nu}=\mathcal K^{\mu
\nu}\frac{\partial S}{\partial x^\mu}\frac{\partial S}{\partial
x^\nu} \, . \label{kilten}
\end{eqnarray}
In both cases, we can use the expression in terms of the $r$
equation or the $u/\theta$ equation. We will choose to work with
the latter in both cases.

For the four dimensional Kerr-Taub-NUT metric analyzed above, the
expression for $K$ from (\ref{sepnut}) is
\begin{eqnarray}
K&=&-\Delta_u\left[\frac{dS_u(u)}{du}\right]^2-\frac{1}{\Delta_u}[u_1u_2E-aL_\phi]^2-m^2u_1u_2
\, .
\end{eqnarray}
Thus, from (\ref{kilten}) we can easily read
\begin{eqnarray}
\mathcal K=-\Delta_u \partial_u \otimes \partial_u -
\frac{1}{\Delta_u}\left[u_1u_2\partial_t\otimes\partial_t + a^2
\partial_\phi \otimes\partial_\phi+sym(au_1u_2\partial_t\otimes\partial_\phi)\right]
\end{eqnarray}

Since this Killing tensor is not a simple linear combination of
Killing vectors, it is non-trivial and irreducible.

For the five dimensional $U(1)^3$ charged metric analyzed above,
the expressions for $K$ from (\ref{sepu1}) is
\begin{eqnarray}
-K&=&\left[\frac{dS_\theta(\theta)}{d\theta}\right]^2+\csc^2\theta
L_\phi ^2 +\sec^2\theta L_\psi ^2 \, .
\end{eqnarray}
Thus, again from (\ref{kilten}) we can read
\begin{eqnarray}
\mathcal
K=-\partial_\theta\otimes\partial_\theta-\frac{1}{\sin^2\theta}
\partial_\phi\otimes\partial_\phi -\frac{1}{\cos
^2\theta}\partial_\psi\otimes\partial_\psi \, .
\end{eqnarray}

This Killing tensor however turns out to be a reducible one. In
this situation, since both rotation parameters, there is an
additional Killing vector which represents the additional symmetry
of being able to rotate each of the two rotation planes into each
other. This Killing tensor can be obtained using linear
combinations of outer products of this Killing vector. Further
details and explicit constructions can be found in \cite{VSP2}.

We can easily check using Maple\cite{Map}, that the Killing
tensors in both spacetimes do satisfy the Killing equation. It is
the existence of these Killing tensors that allows for complete
separation of the Hamilton-Jacobi equation.

\section{The Scalar Field Equation}
Consider a scalar field $\Psi$ in a gravitational background with the action
\begin{equation}
S[\Psi]=-\frac{1}{2}\int d^Dx\sqrt{-g}((\nabla \Psi)^2+ \alpha R \Psi ^2 + m^2
\Psi ^2 ) \,,
\end{equation}
where we have included a curvature dependent coupling. However, in
these Kerr-(Anti) de Sitter backgrounds with charges, $R$ is
constant (proportional to the cosmological constant $\Lambda$). As
a result we can trade off the curvature coupling for a different
mass term. So it is sufficient to study the massive Klein-Gordon
equation in this background. We will simply set $\alpha=0$ in the
following. Variation of the action leads to the Klein-Gordon
equation
\begin{equation}
\frac{1}{\sqrt{-g}}\partial _{\mu}(\sqrt{-g} g^{\mu \nu}\partial _{\nu} \Psi
)=m^2 \Psi \,.\label{KG1}
\end{equation}

\subsection{Massive Scalar Fields in the Four Dimensional
Kerr-Taub-NUT Multicharge Gauged Solution of Supergravity}

Using the explicit expressions for the components of the inverse
metric (\ref{invnut}) and the determinant (\ref{detnut}), the
Klein-Gordon equation for a massive scalar field in this spacetime
can be written as
\begin{eqnarray}
m^2\Psi&=&\frac{1}{W}\left(\frac{1}{\Delta_u \Delta_r
}[\Delta_ru_1^2u_2^2-\Delta_ur_1^2r_2^2]\partial _t ^2 \Psi
+\frac{2a}{\Delta_r\Delta_u}[\Delta_ru_1u_2+\Delta_ur_1r_2]\partial^2_{t\phi}\Psi\right.
\nonumber \\
&+&\left.\frac{2a^2}{\Delta_r\Delta_u}[\Delta_r-\Delta_u]\partial_\phi^2\Psi+\partial_r(\Delta_r\partial_r
\Psi)+\partial_u(\Delta_u \partial_u \Psi) \right)\, .
\end{eqnarray}
(Note that this expression agrees with equation (16) in
\cite{ZGLP} with the notation $p=a\cos\theta$, $q=r$,
$X=\Delta_u$, $Y=\Delta_r$, and $k=0$ in four dimensions for the
uncharged, i.e. $\delta_i=0$, Kerr-Taub-NUT metrics. This is a
good check for consistency.) We assume the usual multiplicative
ansatz for the separation of the Klein-Gordon equation
\begin{eqnarray}
\Psi=\Phi_r(r)\Phi_u(u)e^{-iEt}e^{iL_\phi \phi} \, .
\end{eqnarray}
Then we can easily separate out the $r$ and $u$ dependance as
\begin{eqnarray}
K&=&-\frac{1}{\Phi_u(u)}\frac{d}{du}\left(\Delta_u\frac{d\Phi_u(u)}{du}\right)+m^2u_1u_2+\frac{u_1^2u_2^2}{\Delta_u}E^2-\frac{2au_1u_2}{\Delta_u}EL_\phi+\frac{2a^2L_\phi^2}{\Delta_u}
\, , \nonumber \\
K&=&\frac{1}{\Phi_r(r)}\frac{d}{dr}\left(\Delta_r\frac{d\Phi_r(r)}{dr}\right)-m^2r_1r_2-\frac{r_1^2r_2^2}{\Delta_r}E^2+\frac{2ar_1r_2}{\Delta_r}EL_\phi+\frac{2a^2L_\phi^2}{\Delta_r}
\, ,
\end{eqnarray}
where $K$ is again a  separation constant. At this point we have
completely separated out the Klein-Gordon equation for a massive
scalar field in this spacetime.

\subsection{Massive Scalar Fields in the $U(1)^3$ Gauged
Kerr-(Anti) de Sitter Black Hole Solution of $\mathcal{N} = 2$
Supergravity in Five Dimensions}

Using the explicit expressions for the components of the inverse
metric (\ref{invu1}) and the determinant (\ref{detu1}), the
Klein-Gordon equation for a massive scalar field in this spacetime
can be written as
\begin{eqnarray}
m^2\Psi&=&-\frac{B(r)+R}{r^2W(r)}\partial_t^2\Psi+\frac{2C(r)}{r^2W(r)}(\partial^2_{t\phi}\Psi+\partial_{t\psi}\Psi)+\frac{1}{rR}\partial_r(rRW(r)\partial_r\Psi)
\nonumber \\
&+&\frac{A(r)B(r)\cos^2\theta+A(r)R+C^2(r)\cos^2\theta}{R\sin^2\theta
r^2 W(r)} \partial_\phi ^2 \Psi \nonumber \\
&+&\frac{A(r)B(r)\sin^2\theta+A(r)R+C^2(r)\sin^2\theta}{R\cos^2\theta
r^2 W(r)} \partial_\psi ^2 \Psi
-\frac{2[A(r)B(r)+C^2(r)]}{Rr^2W(r)}\partial^2_{\phi\psi}\Psi
\nonumber \\
&+&\frac{1}{R\sin\theta \cos \theta}\partial_\theta(\sin\theta
\cos\theta \partial_\theta \Psi)\, .
\end{eqnarray}

Again we assume the usual multiplicative ansatz for separation
\begin{eqnarray}
\Psi=\Phi_r(r)\Phi_\theta(\theta)e^{-iEt}e^{iL_\phi
\phi}e^{iL_\psi \psi} \, .
\end{eqnarray}

After extensive algebraic manipulation similar to that of the
Hamilton-Jacobi equation, and the use of some trigonometric
identities along the way, we find that the $r$ and $\theta$
equations decouple into the form
\begin{eqnarray}
K&=&-\frac{1}{\Phi_\theta(\theta)\sin\theta
\cos\theta}\frac{d}{d\theta}\left(\sin\theta \cos \theta
\frac{d\Phi_\theta(\theta)}{d\theta}\right)+\frac{L_\phi^2}{\sin^2\theta}+\frac{L_\psi^2}{\cos^2\theta}
\, , \nonumber \\
K&=&
\frac{1}{\Phi_r(r)r}\frac{d}{dr}\left(rRW(r)\frac{d\Phi_r(r)}{dr}\right)+\frac{B(r)R+R^2}{r^2W(r)}E^2
+ \frac{2C(r)R}{r^2W(r)}(L_\phi+L_\psi) \nonumber \\
&+& \frac{A(r)B(r)+C^2(r)}{r^2W(r)}(L_\phi+L_\psi)^2 \, .
\end{eqnarray}
where $K$ is again a  separation constant. At this point we have
completely separated out the Klein-Gordon equation for a massive
scalar field in this spacetime.

We note the role of the Killing tensors in the separation terms of
the Klein-Gordon equations in both spacetimes. In fact, the
complete integrability of geodesic flow of both metrics via the
Hamilton-Jacobi equation can be viewed as the classical limit of
the statement that the Klein-Gordon equation in both metrics also
completely separates.

\section*{Conclusions}
We studied the complete integrability properties of the
Hamilton-Jacobi and the Klein-Gordon equations in the background
of two recently discovered rotating black hole solutions of
supergravity with charge(s): the four dimensional Kerr-Taub-NUT
Multicharge gauged supergravity solution, and the $U(1)^3$ gauged
Kerr-(Anti) de Sitter black hole solution of $\mathcal{N} = 2$
supergravity in five dimensions. Complete separation of both the
Hamilton-Jacobi and Klein-Gordon equations in these backgrounds in
Boyer-Lindquist-like coordinates is demonstrated. This is due to
the enlarged dynamical symmetry of the spacetime. We construct the
Killing tensors (one of them irreducible) in both spacetimes which
explicitly permits complete separation. We also derive first-order
equations of motion for classical particles in these backgrounds,
and analyze the properties of some special trajectories. It should
be emphasized that these complete integrability properties are a
fairly non-trivial consequence of the specific form of the
metrics, and generalize several such remarkable properties for
other previously known metrics.

Further work in this direction could include the study of
higher-spin field equations in these backgrounds, which is of
great interest, particularly in the context of string theory.
Explicit numerical study of the equations of motion for specific
values of the black hole parameters could lead to interesting
results. The geodesic equations presented can also readily be used
in the study of black hole singularity structure in an AdS
background using the AdS/CFT correspondence.

\end{document}